\documentclass{elsart}

\usepackage[square,comma]{natbib}
\usepackage{graphicx}
\usepackage{pxfonts}
\usepackage{lineno}

\usepackage{amssymb}

\journal{}

\begin{document}
\thispagestyle{empty}
\begin{Large}
\textbf{DEUTSCHES ELEKTRONEN-SYNCHROTRON}

\textbf{\large{Ein Forschungszentrum der
Helmholtz-Gemeinschaft}\\}
\end{Large}

DESY 11-009

January 2011

\begin{eqnarray}
\nonumber &&\cr \nonumber && \cr \nonumber &&\cr
\end{eqnarray}
\begin{eqnarray}
\nonumber
\end{eqnarray}
\begin{center}
\begin{Large}
\textbf{Improvement of the crossed undulator design for effective circular polarization
control in X-ray FELs}
\end{Large}
\begin{eqnarray}
\nonumber &&\cr \nonumber && \cr
\end{eqnarray}

\begin{large}
Gianluca Geloni,
\end{large}
\textsl{\\European XFEL GmbH, Hamburg}
\begin{large}

Vitali Kocharyan and Evgeni Saldin
\end{large}
\textsl{\\Deutsches Elektronen-Synchrotron DESY, Hamburg}
\begin{eqnarray}
\nonumber
\end{eqnarray}
\begin{eqnarray}
\nonumber
\end{eqnarray}
ISSN 0418-9833
\begin{eqnarray}
\nonumber
\end{eqnarray}
\begin{large}
\textbf{NOTKESTRASSE 85 - 22607 HAMBURG}
\end{large}
\end{center}
\clearpage
\newpage

\begin{frontmatter}



\title{Improvement of the crossed undulator design for effective circular polarization
control in X-ray FELs}


\author[XFEL]{Gianluca Geloni\thanksref{corr},}
\thanks[corr]{Corresponding Author. E-mail address: gianluca.geloni@xfel.eu}
\author[DESY]{Vitali Kocharyan}
\author[DESY]{and Evgeni Saldin}

\address[XFEL]{European XFEL GmbH, Hamburg, Germany}
\address[DESY]{Deutsches Elektronen-Synchrotron (DESY), Hamburg,
Germany}

\begin{abstract}
The production of X-ray radiation with a high degree of circular polarization constitutes an important goal at XFEL facilities. A simple scheme to obtain circular polarization control with crossed undulators has been proposed so far. In its simplest configuration the crossed undulators consist of pair of short planar undulators in crossed position separated by  an electromagnetic phase shifter.  An advantage of this configuration is a fast helicity switching. A drawback is that a high degree  of circular polarization (over 90 \%) can only be achieved for lengths of the insertion devices significantly shorter than the gain length, i.e. at output power significantly lower than the saturation power level. The obvious and technically possible extension considered in this paper, is to use a setup with two or more crossed undulators separated by phase shifters. This cascade crossed undulator scheme is distinguished, in performance, by a fast helicity switching, a high degree of circular polarization (over 95 \%) and a high output power level, comparable with the saturation power level in the baseline undulator at fundamental wavelength. We present feasibility study and exemplifications for the LCLS baseline in the soft X-ray regime.
\end{abstract}

%
%
%
\end{frontmatter}



\section{\label{sec:intro} Introduction}

\begin{figure}
\includegraphics[width=1.0\textwidth]{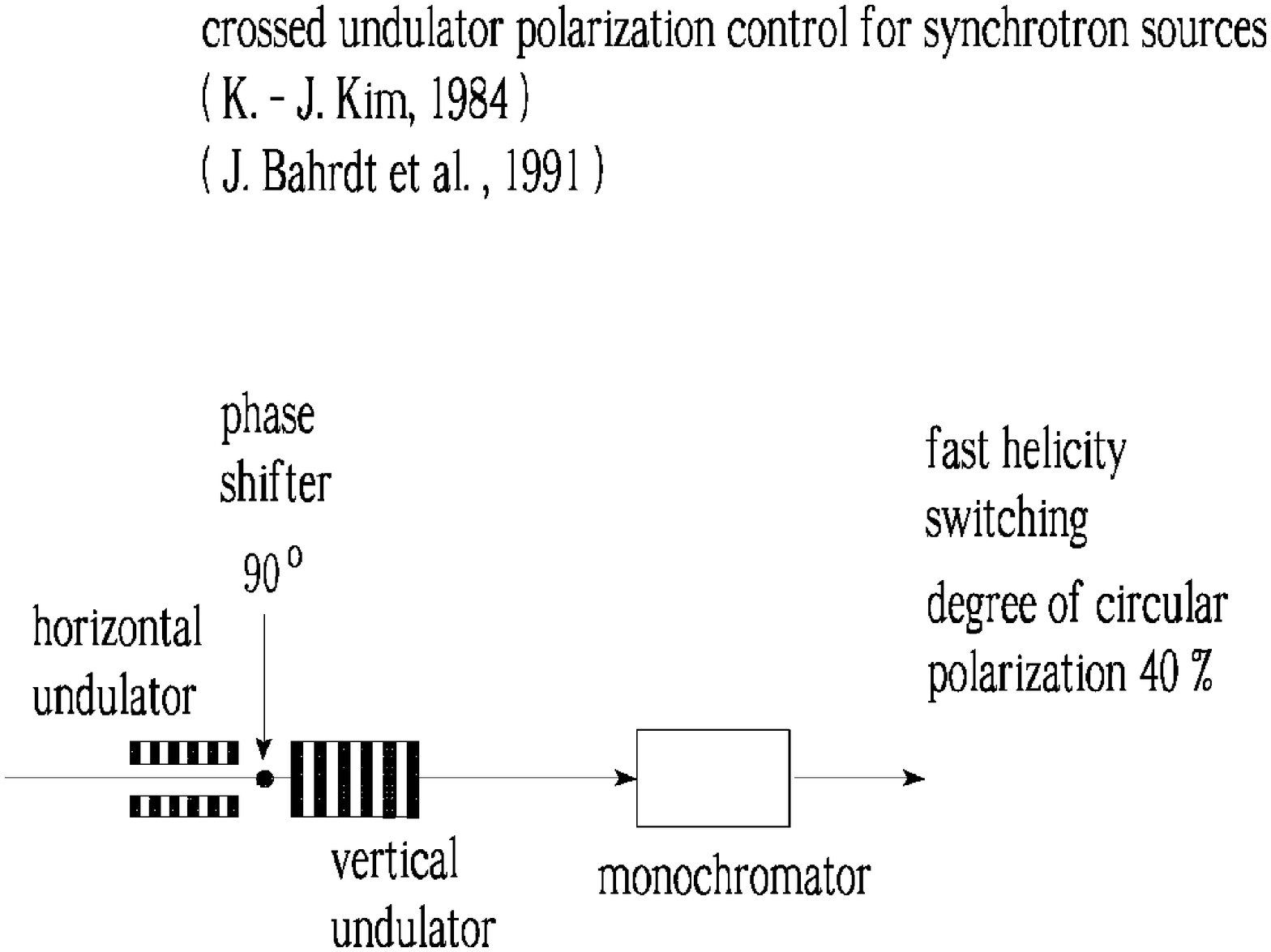}
\caption{Scheme with crossed undulator for synchrotron radiation sources} \label{lclspcms7}
\end{figure}

\begin{figure}
\includegraphics[width=1.0\textwidth]{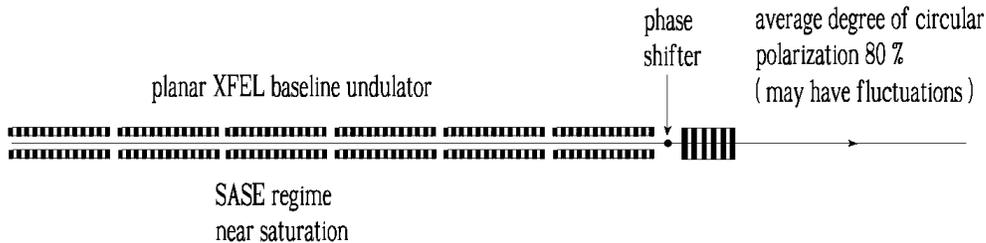}
\caption{Scheme with crossed undulator for X-ray FEL sources} \label{lclspcms6}
\end{figure}

\begin{figure}
\includegraphics[width=1.0\textwidth]{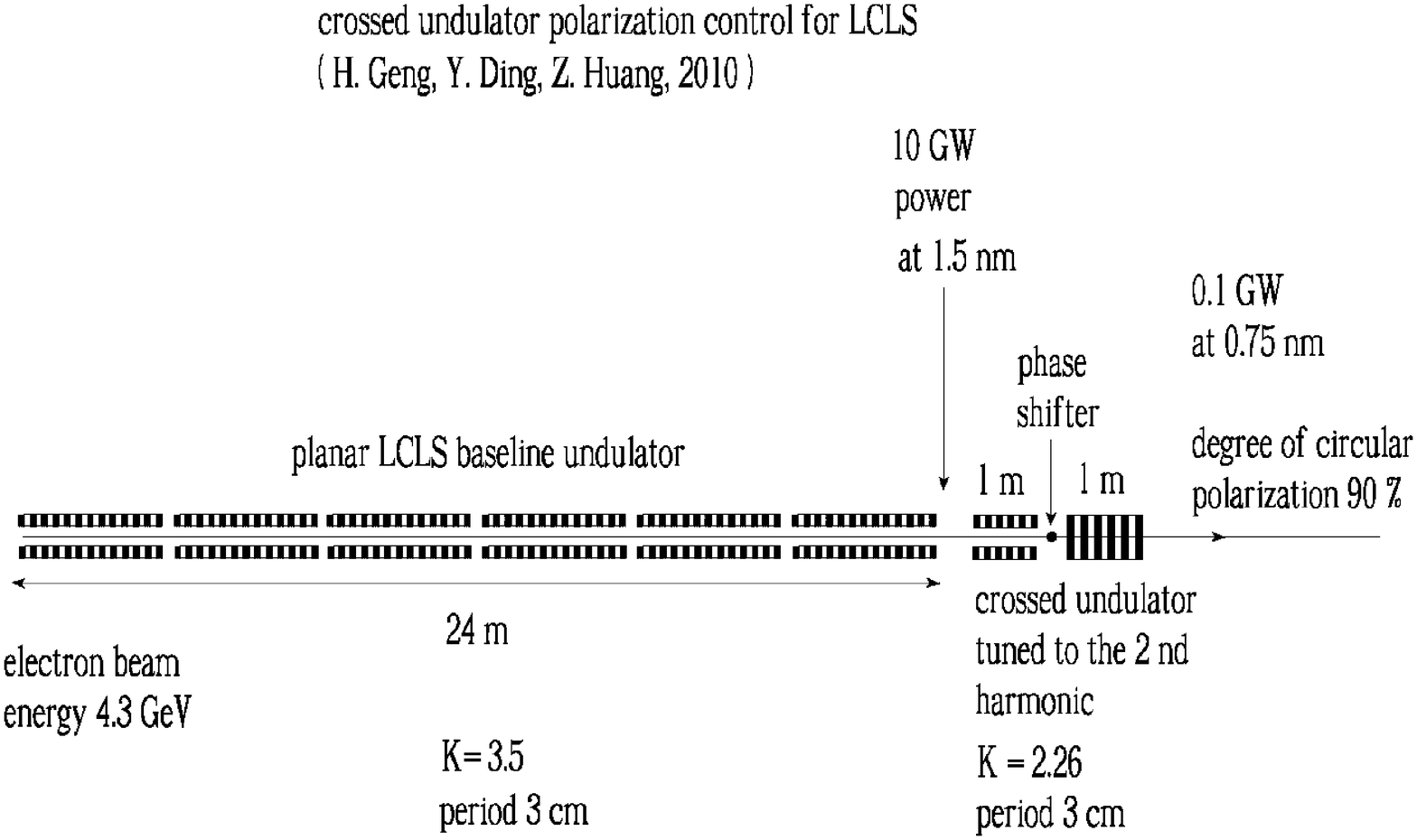}
\caption{Schematic of the crossed undulator proposed for polarization control at the LCLS} \label{lclspcms5}
\end{figure}

\begin{figure}
\includegraphics[width=1.0\textwidth]{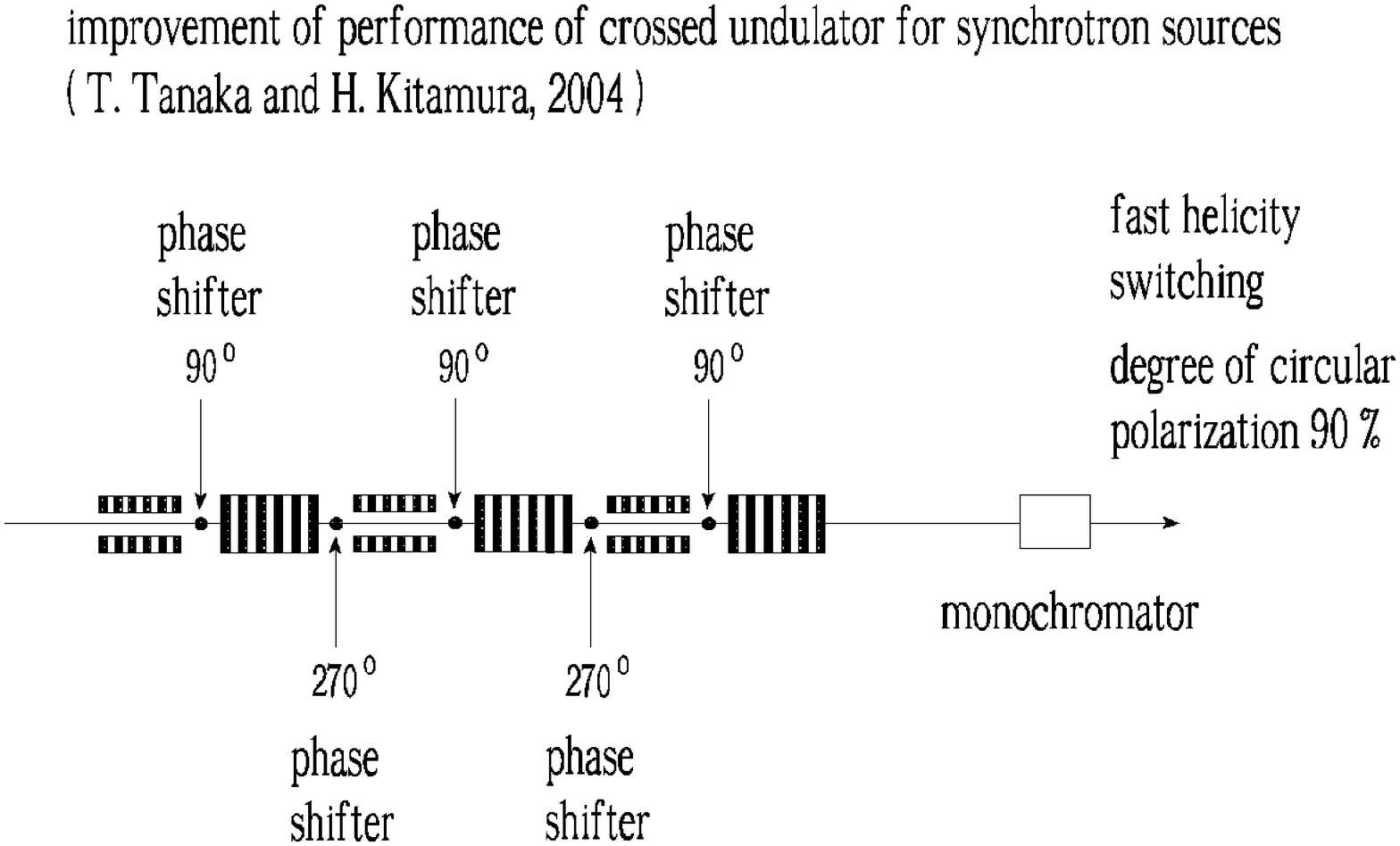}
\caption{Schematic of the cascade crossed undulator proposed for effective polarization control at synchrotron radiation sources} \label{lclspcms2}
\end{figure}

The LCLS baseline includes a planar undulator system, which produces horizontally polarized X-ray pulses \cite{LCLS2}. However, there is an always increasing demand by the LCLS users for circularly polarized X-ray pulses with fast switching of helicity. An APPLE-type undulator \cite{SASA} can provide variable polarization, but it is difficult to quickly change the polarity due to the magnet motion mechanism and, thus, to achieve fast helicity switching. One possible solution is to use crossed undulators. The concept of crossed undulators was devised by Kim in order to produce various polarization states with planar undulators \cite{KIM1}. The configuration is based on a pair of planar undulators in a crossed position, Fig. \ref{lclspcms7}. A phase shifter between the undulators controls the radiation phase between horizontal and vertical components by providing a bump orbit to the electron beam with its magnetic field. This allows for the generation of various polarization states such as circular ones. A fast switching of the polarization direction (up to the kHz level) can be achieved by using electromagnetic phase shifters. When this concept is applied to synchrotron radiation sources, the radiation pulses generated in horizontal and vertical undulators by each electron do not overlap in time. Thus, a monochromator after the crossed undulator is required to temporally stretch both pulses and to achieve interference, Fig. \ref{lclspcms7}.  The crossed undulator setup has been installed at BESSY and the degree of circular polarization was found to be in the range of $40-45 \%$ \cite{BAH1}.  The degree of polarization is limited, in practice, by the finite beam emittance, energy spread and resolution of the monochromator. In particular, the angular divergence of the electron beam is responsible for a blurring of the phase between the radiation field components, which is a cause of depolarization. A remarkable feature of an XFEL device is a narrow bandwidth in the order of  $0.1\%$ of the output radiation. It follows that the monochromator, which is needed for the operation of the crossed undulator scheme at synchrotron radiation sources, can be avoided. In addition, due to the high quality of the electron beam at XFELs, emittance and energy spread effects play no significant roles in the determination of the degree of polarization for crossed undulators.

Three different approaches have been proposed so far for the production of circularly polarized radiation with crossed undulators at XFELs facilities. A first approach was proposed in \cite{KIMI} and further studied in \cite{DING}. In this scheme, a short planar undulator is placed behind the long planar undulator, oriented orthogonally to it, Fig. \ref{lclspcms6}.  If the length of the short undulator is approximately 1.3 times the FEL gain length, the two orthogonal linear components have equal intensities. Therefore, if their phase difference is $\pi/2$, their combination results in circular polarization. According to simulations \cite{DING}, the maximum degree of circular polarization in the regime of SASE saturation is over $80 \%$.  The 3D FEL gain length, however, depends on a number parameters such as wavelength, peak current, emittance, energy spread, beta function. Some of them might fluctuate leading to fluctuations of the polarization degree as well. In addition, this scheme can only be optimized for one wavelength with fixed parameters, while for other parameter choices the degree of polarization drops.  A modified second scheme has been proposed in \cite{GENG} in order to improve stability, Fig. \ref{lclspcms5}. The planar baseline undulator is only used as a buncher, while a short pair of crossed planar undulators tuned to the second harmonic is placed behind the buncher. The radiation in the baseline undulator is characterized by a different frequency than that produced in the crossed undulator, and therefore has no effect on the polarization properties of the harmonic fields. Moreover, the second harmonic contents generated through the nonlinear harmonic generation process in the baseline undulator are smaller than those generated in the crossed undulator, and can be ignored. The maximum degree of circular polarization achievable in this case is over $90 \%$ at $0.1$ GW power level for the LCLS \cite{GENG}.  Finally, a third scheme where the crossed undulator configuration works at the fundamental wavelength has been proposed in \cite{YLI}. However, in order to practically exploit that method, the electron beam needs to be bent in order to separate the radiation generated in the crossed undulator. This poses the non-trivial challenge of preserving the electron beam microbunching \cite{LI1}.

The drawback of crossed undulator XFEL schemes is that, in order to reach a circular degree of polarization larger than $80-90 \%$, undulators
need to be significantly shorter than the FEL gain length. This is due to the need for equal intensities in the two linearly polarized components separately generated in the different undulators. As a result, the performance of the output radiation is significantly lower than that of the light produced by the baseline undulator, meaning that the intensity is reduced by more than an order of magnitude.

The cascade crossed undulator scheme proposed in \cite{TANA}, Fig. \ref{lclspcms2} for synchrotron radiation sources is a candidate to overcome this difficulty. In this paper, we study the use of this scheme at XFEL facilities as a mean to generate circularly polarized radiation  at the fundamental frequency.  The undulator is composed of several cascades, each of which forms a crossed undulator. We present exemplifications
for the LCLS baseline case.  The radiation from the proposed device is investigated numerically, and shows that a high degree (over $95 \%$) of circular polarization and, simultaneously, a high output power level (10 GW-level) can be obtained if a sufficiently large number of cascades (up to four) is considered.

The applicability of the cascade crossed undulator scheme is obviously not restricted to the LCLS baseline. Other facilities, e.g. the LCLS-II and the European XFEL, may benefit from this scheme as well.

\section{\label{sec:scheme} Possible circular polarization control scheme with cascade
crossed undulators for the LCLS baseline}

\begin{figure}
\includegraphics[width=1.0\textwidth]{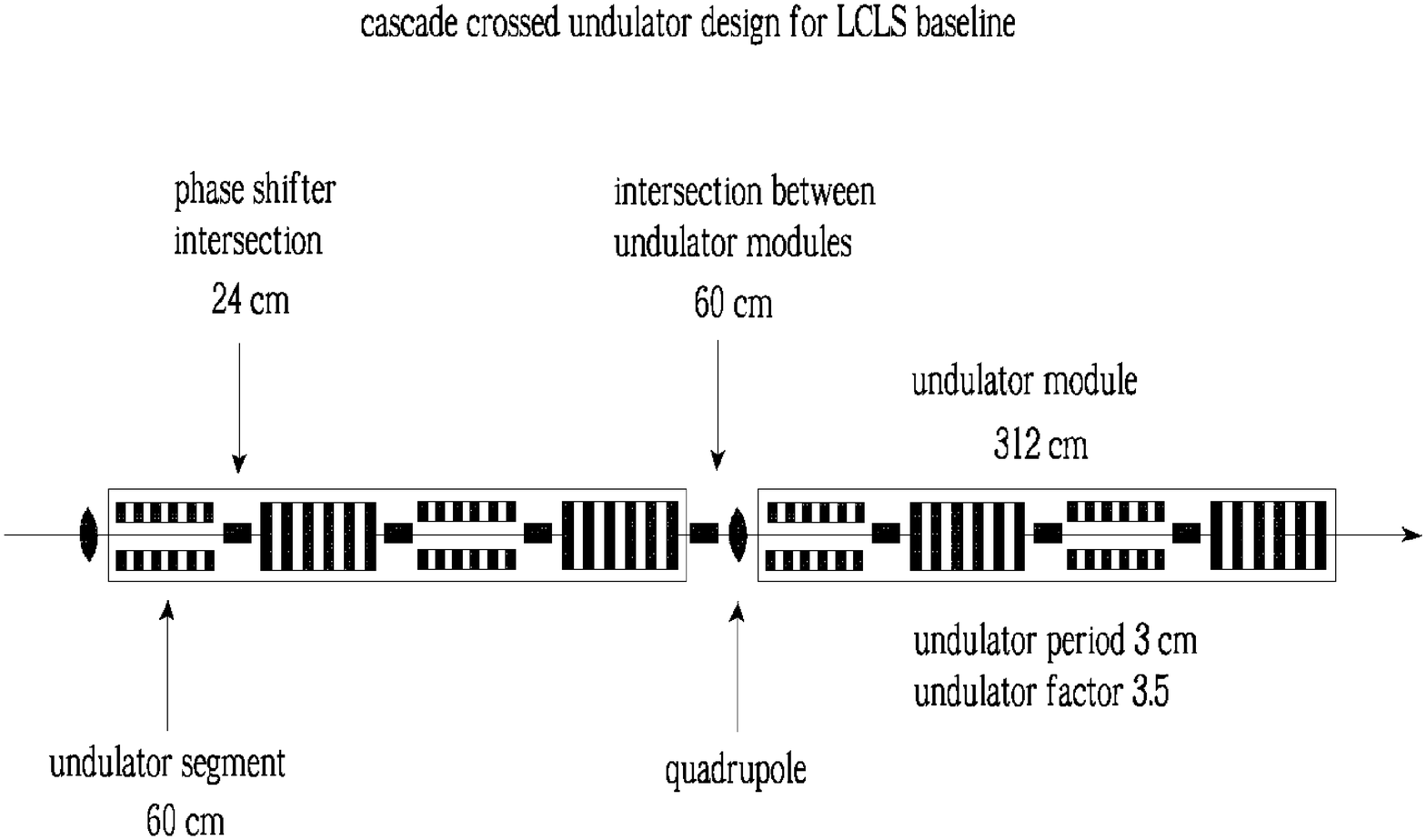}
\caption{Schematic of the cascade crossed undulator proposed for effective polarization control at the LCLS baseline.} \label{lclspcms3}
\end{figure}

\begin{figure}
\includegraphics[width=1.0\textwidth]{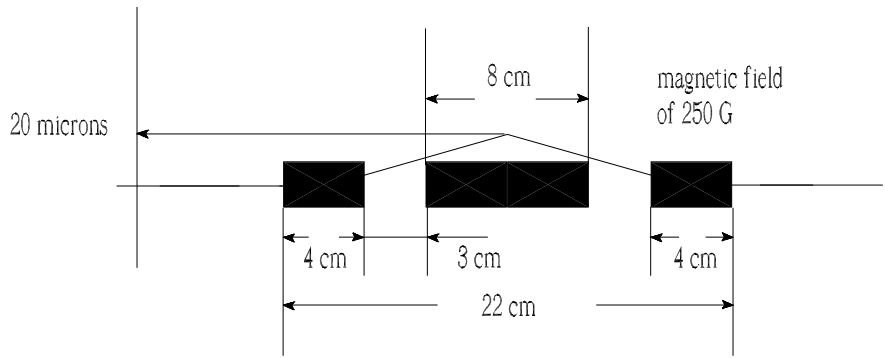}
\caption{Phase shifter design for the cascade crossed undulator at LCLS. Each phase shifter controls the radiation phase between undulator segments by providing a bump orbit to the electron beam with its magnetic field in order to generate various polarization states.} \label{lclspcms8}
\end{figure}
\begin{figure}
\includegraphics[width=1.0\textwidth]{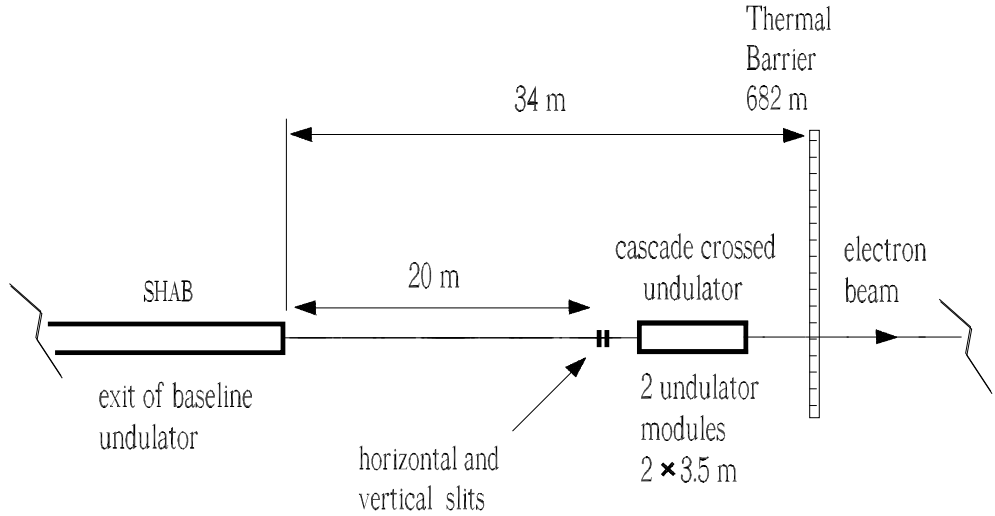}
\caption{The installation of the cascade crossed undulator after the LCLS baseline undulator.} \label{lclspcms1a}
\end{figure}

\begin{figure}
\includegraphics[width=1.0\textwidth]{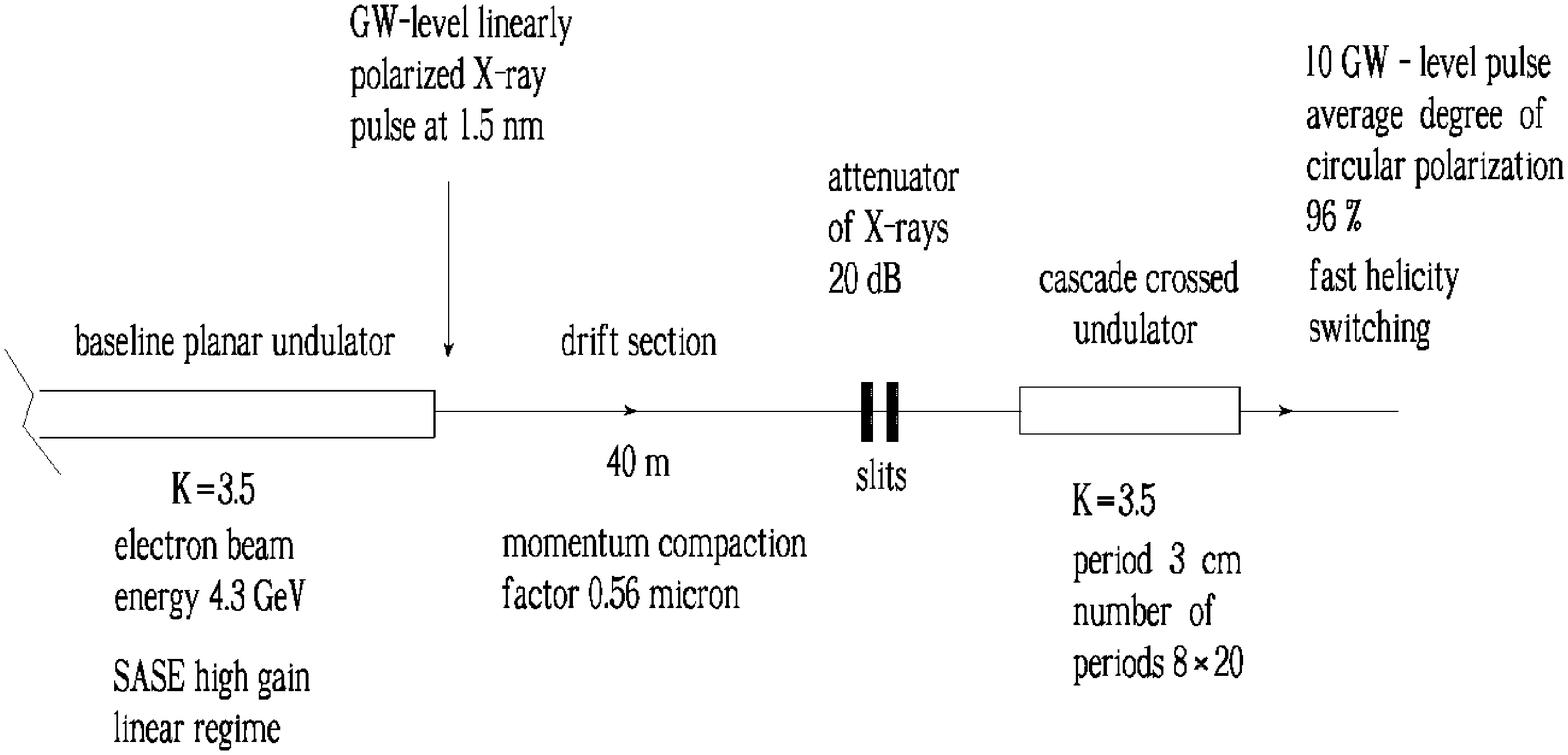}
\caption{Design of the undulator system for effective polarization control at
the LCLS baseline.} \label{lclspcms4}
\end{figure}
The subject of this article is the proposal of an undulator configuration for X-ray FELs allowing for high and stable degree of circular polarization, high output power and fast helicity switching. Although it is based on the short crossed undulator configuration, this configuration can achieve much higher output power. With reference to Fig. \ref{lclspcms3}, the undulator is composed of several cascades, each of which forms a crossed undulator. The helicity switching can be performed very quickly \cite{NAKA}. In fact, a fast switching of the polarization direction can be achieved by using electromagnetic phase shifters, Fig. \ref{lclspcms8}.

In this section we describe a particular realization of our proposal, a polarization control scheme that may be easily developed at the LCLS. It combines the cascade crossed undulator arrangement proposed in \cite{TANA} with the filtering method proposed in \cite{OURC}. An overall sketch of the setup is shown in Fig. \ref{lclspcms1a}. The electron beam first goes through the baseline undulator, producing SASE radiation. This induces energy and density modulation on the electron beam. Following \cite{OURC} we assume that the five second harmonic afterburner (SHAB) modules are rolled away \cite{BAIL} from the beamline, Fig. \ref{lclspcms4}. In this way, we provide a total of $40$ m straight section for the electron beam, $20$ m corresponding to the SHAB modules, and further $20$ m corresponding to the straight section after the exit of the main undulator. At the end of the $40$ m-long straight section we place horizontal and vertical slits. The scheme is similar to that presented in \cite{OURC}, with the difference that after the slits we install a cascade crossed undulator, instead of an APPLE-type undulator.

After the straight section, electron beam and radiation pass through horizontal and vertical slits, suppressing the linearly-polarized soft X-ray radiation from the LCLS baseline undulator. Since the slits are positioned 40 m downstream of the planar undulator, the radiation pulse has a ten times larger spot size compared with the electron bunch transverse size, and the background radiation power can therefore be diminished of two orders of magnitude. As discussed in \cite{OURC}, the slits can be made of Beryllium foils, for a total thickness of $150~\mu$m.  Such foils will block the radiation, but will let the electrons go through \cite{EMMA}. The advantage of the spoiling scheme is that radiation is attenuated of $20$ dB, while the halo of the electron bunch is allowed to propagate through the setup up to the beam dump without electron losses. Ionization losses can be neglected.

As already remarked in \cite{OURC}, one should account for the fact that the straight section acts as a dispersive element. Therefore, a klystron-like bunching effect should also be accounted for, which modifies the density modulation from that found at the exit of the first undulator. From this viewpoint, the first (baseline) LCLS undulator behaves as an energy modulator, and the drift section, i.e. the straight section, transforms the energy into density modulation. Estimations for the klystron-bunching effect and the influence of betatron motion have already been presented in \cite{OURC}. Here we only sum up the conclusions reached in that reference, repeating that these effects are not preventing the scheme from working, but that the planar baseline undulator should be operated in the linear regime.

Following the slits, the electron beam enters the crossed-undulator cascade, where the microbunched electron beam produces intense bursts of radiation in any selected polarization state.  For simplicity we assume average betatron function values $\beta = 10$ m at $1.5$ nm, equal to the betatron function in the baseline, since it is advantageous, for feasibility study purposes to assume that there is no significant difference in the betatron function along the $20$ m-long drift section and in the SHAB undulator focusing system. Then, we propose to fill two undulator modules of about $3.5$ m of length each with crossed-undulator cascades as sketched in Fig. \ref{lclspcms3}. We assume that each undulator segment is $60$ cm-long, with a period of $3$ cm. This allows for the installation of phase shifters between one segment and the following one, according to the scheme in Fig. \ref{lclspcms8}. A detailed study concerning the phase shifters can be found in \cite{NAKA}. Two cascades fit in a single module.

\section{\label{sec:sims} FEL simulations}

In the previous Section we gave a qualitative description of the scheme for polarization control. Here we present more detailed FEL simulations with the help of the FEL code GENESIS 1.3 \cite{GENE} running on a parallel machine. We present a statistical analysis consisting of $100$ runs. Parameters used in the simulations for the low-charge mode of operation in the crossed undulators are presented in Table \ref{tt1}. The choice of the low-charge mode of operation is motivated by simplicity.

\begin{table}
\caption{Parameters for the low-charge mode of operation at LCLS used in
this paper.}

\begin{small}\begin{tabular}{ l c c}
\hline & ~ Units &  ~ \\ \hline
Undulator period      & mm                  & 30     \\
K parameter (rms)     & -                   & 3.5  \\
Wavelength            & nm                  & 1.5   \\
Energy                & GeV                 & 4.3   \\
Charge                & nC                  & 0.02 \\
Bunch length (rms)    & $\mu$m              & 1    \\
Normalized emittance  & mm~mrad             & 0.4    \\
Energy spread         & MeV                 & 1.5   \\
\hline
\end{tabular}\end{small}
\label{tt1}
\end{table}

\begin{figure}
\includegraphics[width=1.0\textwidth]{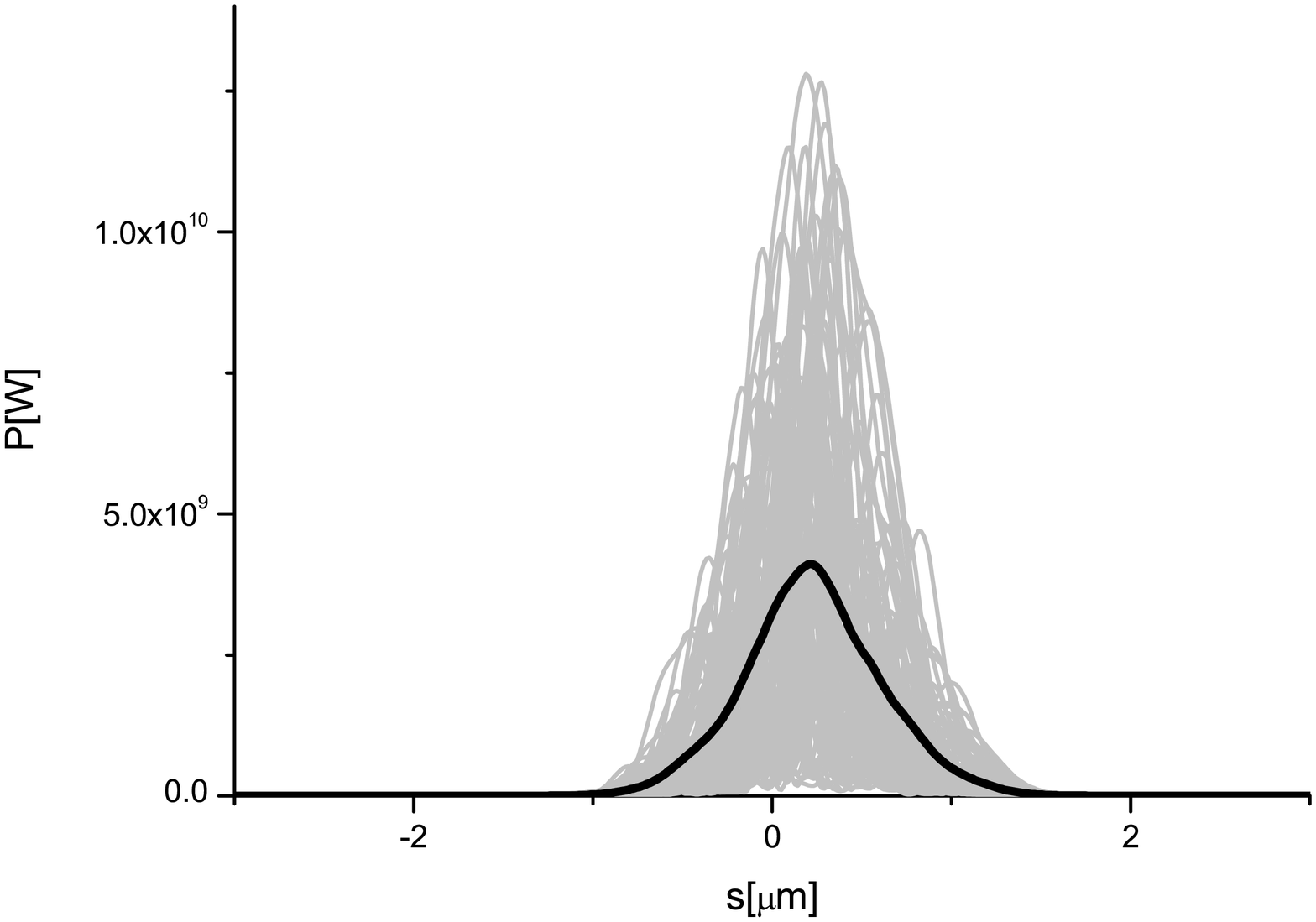}
\caption{Power distribution after the first SASE undulator (5 cells). Grey lines refer to single shot realizations, the black line refers to the average over a hundred realizations. } \label{2power0}
\end{figure}
\begin{figure}
\includegraphics[width=1.0\textwidth]{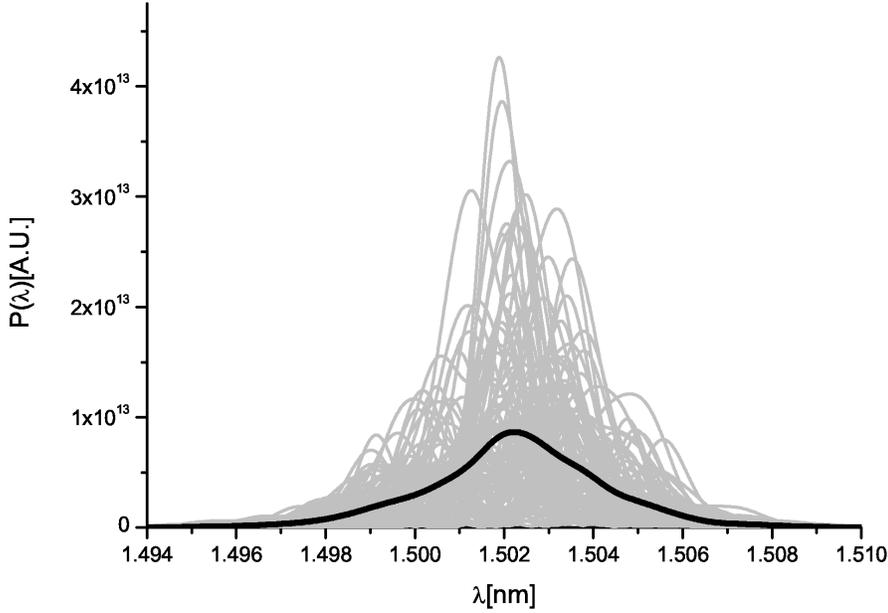}
\caption{Spectrum after the first SASE undulator (5 cells). Grey lines refer to single shot realizations, the black line refers to the average over a hundred realizations.} \label{1spectr0}
\end{figure}

First, the LCLS baseline is simulated. As discussed before, and already considered in \cite{OURC}, the baseline LCLS undulator should work in the linear regime. An optimum is found when only the last $5$ cells upstream of the SHAB are used. In other words we assume that first 23 baseline undulator modules are detuned. The power and spectrum after the baseline undulator are shown in Fig. \ref{2power0} and Fig. \ref{1spectr0}.

The particle file produced by Genesis at the exit of the baseline undulator is downloaded and transformed assuming a dispersive element with $R_{56} \simeq L/\gamma^2 \simeq 560$ nm, which models the following $40$-m long straight section. The transformed particle file is used as an input for further simulations through the setup described in the previous Section, Fig. \ref{lclspcms3}. The average betatron function is assumed to be $\beta = 10$ m. By this we assume that the same focusing system in the SHAB section is continued through the following $20$ m-long straight section and into the cascade cross-undulator. Such assumption can obviously be relaxed, and it is considered here for simplicity reasons only. Also, the influence of the betatron motion on the microbunching is only estimated in \cite{OURC}, but is not explicitly accounted for in simulations. However, from those estimations we expect that such influence would be even smaller than the influence due to the finite energy spread of the beam.

\begin{figure}
\includegraphics[width=1.0\textwidth]{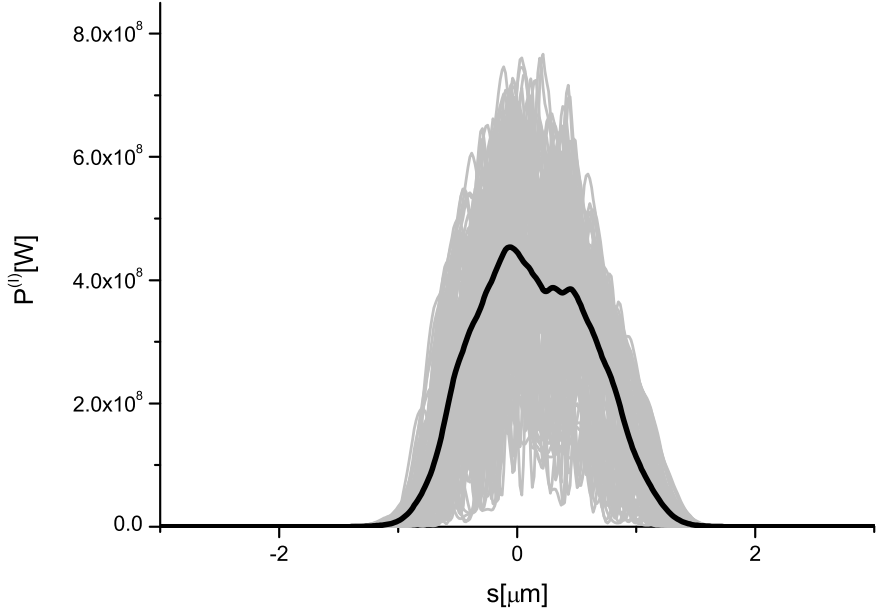}
\caption{Power after the first crossed-undulator cascade, composed by the first crossed undulator. Grey lines refer to single shot realizations, the black line refers to the average over a hundred realizations.} \label{4Pfirst}
\end{figure}
\begin{figure}
\includegraphics[width=1.0\textwidth]{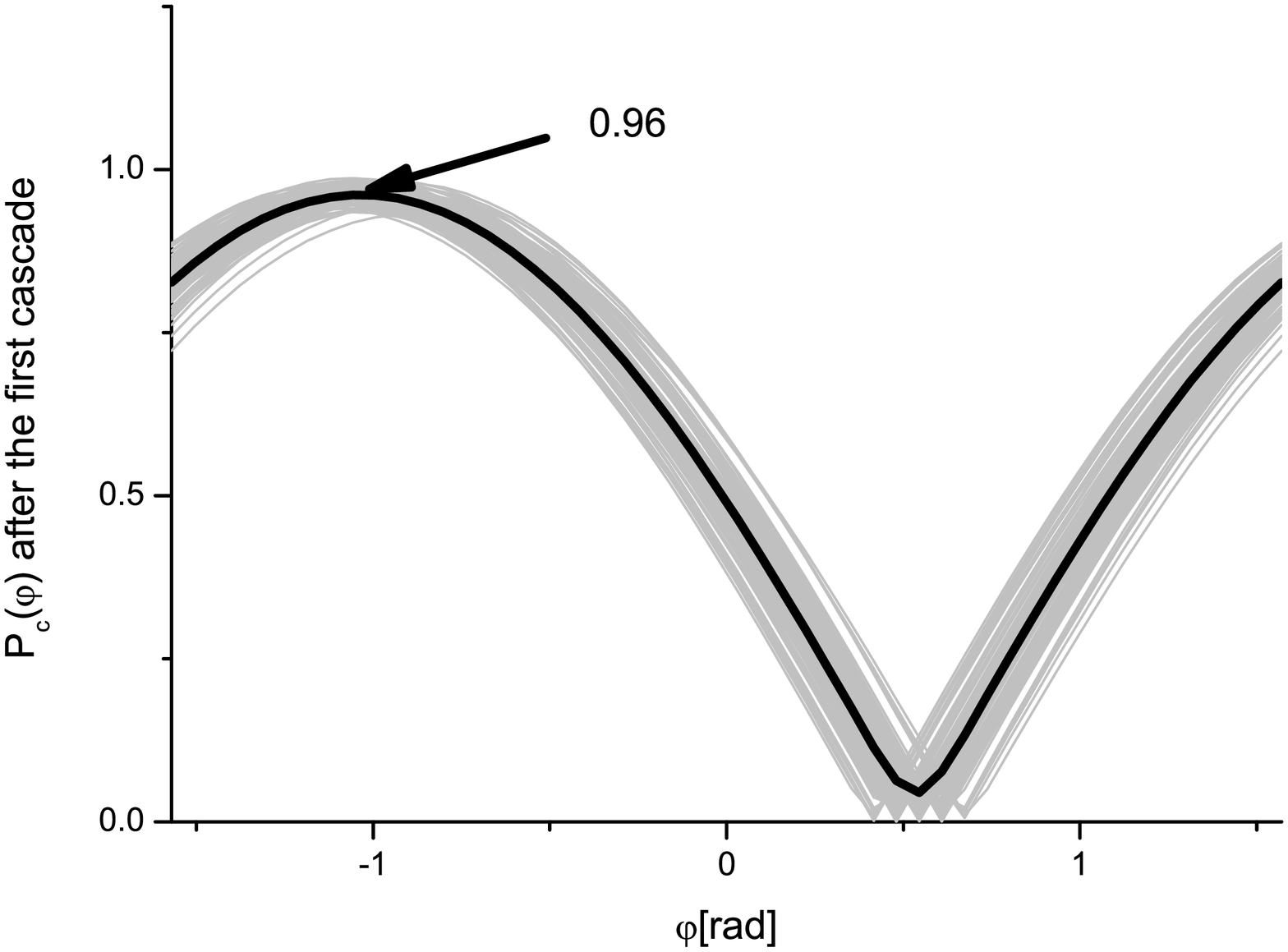}
\caption{Circular degree of polarization as a function of the phase shift, after the first crossed-undulator cascade. Grey lines refer to single shot realizations, the black line refers to the average over a hundred realizations. An average $96\%$ degree of polarization can be obtained when the right phase is chosen.} \label{3Pcphifirst}
\end{figure}
The particle file is used to simulate the radiation output from the first undulator in the cascade. In particular, both a field file and a new particle file are produced by Genesis. The intersection between the first and the second undulator is modeled as a straight section. The correct phase shift is simulated by properly adjusting the length of this straight section. The horizontally polarized radiation should be further propagated through the second undulator and through the second phase shifter. To this purpose we still used Genesis after switching off the interaction with the electron bunch\footnote{This is done by decreasing the electron bunch current to $1$A.}. Then, the particle file at the entrance of the second undulator is used as an input for Genesis to calculate the vertically polarized field from the second undulator.  The total output power from the first cascade, composed by the first crossed undulator is shown in Fig. \ref{4Pfirst}.

In order to calculate the average degree of polarization we used the simplified approach presented in \cite{DING}. Instead of averaging over a three-dimensional field distribution we reduce the averaging procedure to a one-dimensional calculation by, first, taking the Fourier transform of the horizontal and vertical radiation field at this position down the setup. This yields the far-zone radiation field. The on-axis far-zone field is then used to calculate the Stokes parameters and yields the circular polarization degree as a function of time for a given pulse, $P_{1c}(t)$. We subsequently weight $P_{1c}(t)$ over the on-axis power density of the pulse, $I(t)$, and we make an ensemble average over many pulses according to

\begin{eqnarray}
P_c = \frac{1}{N_p} \sum_{n=1}^{N_p}\frac{ \int_{-\infty}^{\infty} I(t) P_{1c}(t) dt}{\int_{-\infty}^{\infty} I(t)  dt} ~,
\label{Pc}
\end{eqnarray}
$N_p = 100$ being the number of pulses in our statistical run. The degree of polarization is obviously a function of the phase between the two polarization components. A plot of the circular degree of polarization as a function of the phase shift, after the first crossed-undulator cascade is shown in Fig. \ref{3Pcphifirst}. Grey lines refer to single shot realizations, the black line refers to the average over a hundred realizations. An average $96\%$ degree of polarization can be obtained when the right phase is chosen.

\begin{figure}
\includegraphics[width=1.0\textwidth]{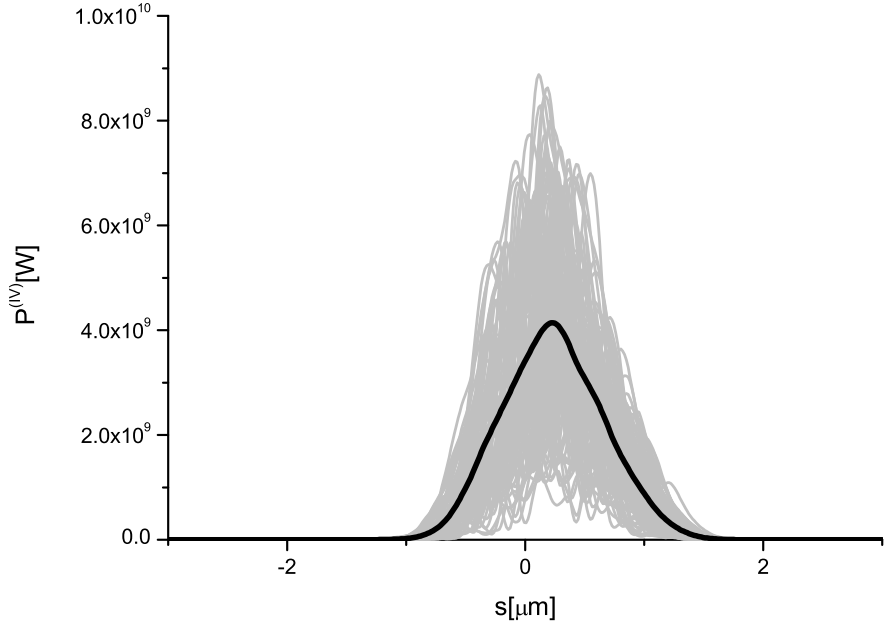}
\caption{Power after the fourth crossed-undulator cascade. Grey lines refer to single shot realizations, the black line refers to the average over a hundred realizations.} \label{6Pfour}
\end{figure}

\begin{figure}
\includegraphics[width=1.0\textwidth]{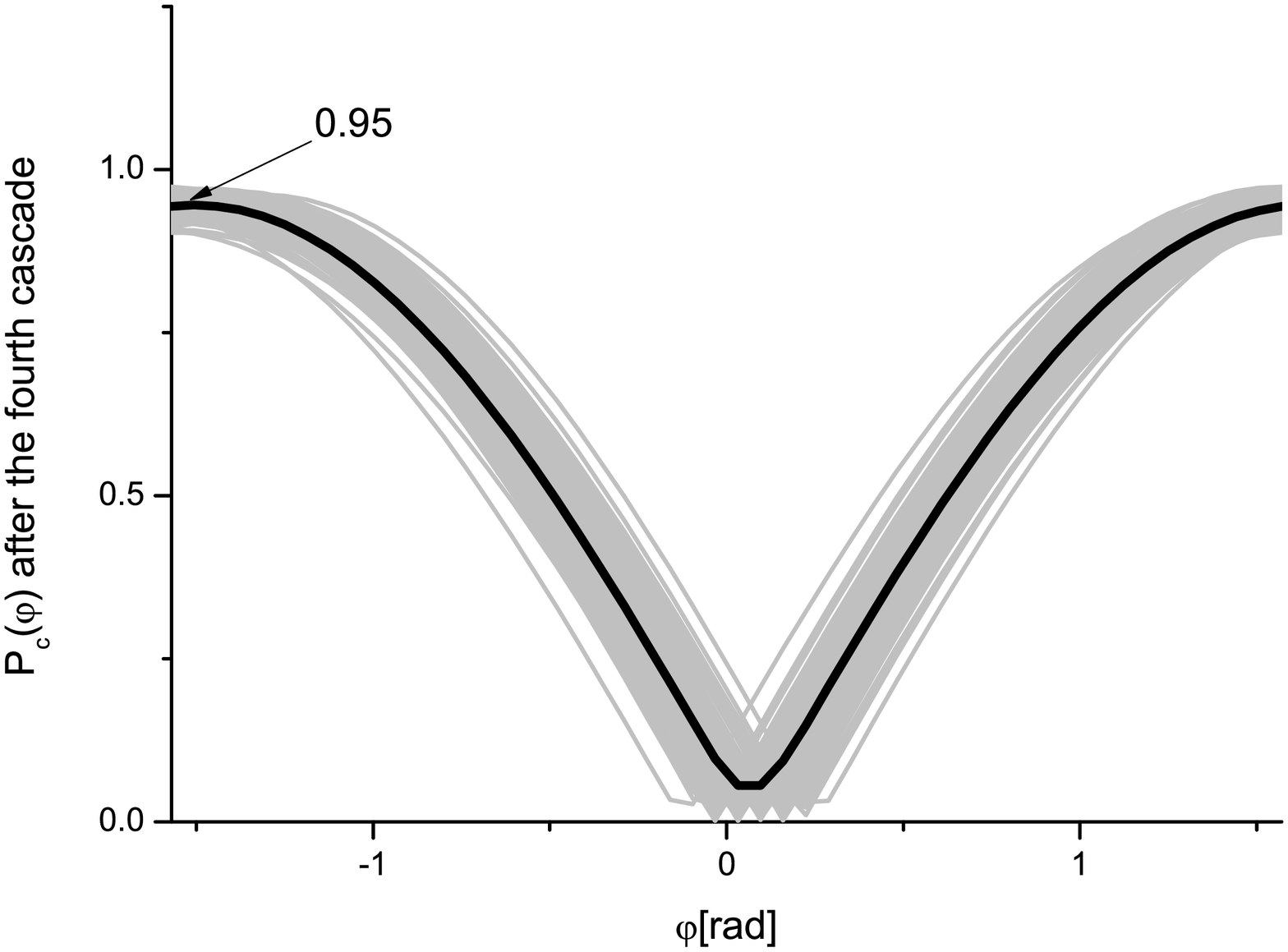}
\caption{Circular degree of polarization as a function of the phase shift, after the fourth crossed-undulator cascade. Grey lines refer to single shot realizations, the black line refers to the average over a hundred realizations. An average $95\%$ degree of polarization can be obtained when the right phase is chosen.} \label{5Pcphifour}
\end{figure}
The particle file resulting from the propagation at nominal current and the field file for the horizontally and vertically polarized radiation are used as input files for the following cascade, and the process is repeated up to the last cascade. The total output power is shown in Fig. \ref{6Pfour}, while the degree of polarization as a function of the phase is shown in Fig. \ref{5Pcphifour}. An average $95\%$ degree of polarization can be obtained when the right phase is chosen, and high-power pulses can be produced in the $10$ GW-level.

\section{Conclusions}

In this paper we exploit the cross-undulator cascade scheme developed in \cite{TANA} in order to achieve ultimate performance in polarization control, yielding  high-power  pulses of X-ray radiation with arbitrary state of polarization, very high degree of polarization, and fast switching of helicity. The proposed setup can be composed of an unlimited number of cascades, up to the full scale of the baseline undulator, the degree of polarization being independent of the length of the setup. This hints to possible future designs for baseline XFEL undulators where, for example, one half of the total length can be taken by a long cross-undulator cascade. In this case one may achieve circular polarization for soft and hard X-rays using the same scheme, and without problems of linearly polarized radiation background from the first part of the undulator.

We presented an illustration of the scheme for the LCLS,  although other facilities like the LCLS-II and the European XFEL may also benefit from it, limiting ourselves to the soft x-ray range. We combined this method with the filtering concept considered in \cite{OURC}. The main advantage achieved consists in obtaining fast helicity-switching up the the KHz level and, simultaneously, in retaining the $10$-GW power level typical of APPLE-type undulators. In fact, by increasing the number of cascades from one to four we increase the output power by ten times from half GW  to $5$ GW at the same high ($95 \%$) degree of polarization. Moreover, the use of planar devices is much less expensive compared to the APPLE-type. The exploitation of the filtering concept, which was first presented in the case of an APPLE-type radiator \cite{OURC}, solves the issue of separating planar from circular polarization radiation components. The setup can be installed at the LCLS in a little time. It constitutes a cost-effective, risk-free alternative to currently available methods for polarization control.

\section{Acknowledgements}

We are grateful to Massimo Altarelli, Reinhard Brinkmann, Serguei
Molodtsov and Edgar Weckert for their support and their interest
during the compilation of this work.

\end{document}